# Stress channelling in extreme couple-stress materials Part I: Strong ellipticity, wave propagation, ellipticity, and discontinuity relations

Panos A. Gourgiotis - Davide Bigoni[1]
email: p.gourgiotis@unitn.it, bigoni@ing.unitn.it
University of Trento, via Mesiano 77, I-38123 Trento, Italy

**Abstract.** Materials with extreme mechanical anisotropy are designed to work near a material instability threshold where they display stress channelling and strain localization, effects that can be exploited in several technologies. Extreme couple stress solids are introduced and for the first time systematically analyzed in terms of several material instability criteria: positive-definiteness of the strain energy (implying uniqueness of the mixed b.v.p.), strong ellipticity (implying uniqueness of the b.v.p. with prescribed kinematics on the whole boundary), plane wave propagation, ellipticity, and the emergence of discontinuity surfaces. Several new and unexpected features are highlighted: (i.) Ellipticity is mainly dictated by the 'Cosserat part' of the elasticity and (ii.) its failure is shown to be related to the emergence of discontinuity surfaces; (iii.) Ellipticity and wave propagation are not interdependent conditions (so that it is possible for waves not to propagate when the material is still in the elliptic range and, in very special cases, for waves to propagate when ellipticity does not hold). The proof that loss of ellipticity induces stress channelling, folding and faulting of an elastic Cosserat continuum (and the related derivation of the infinite-body Green's function under antiplane strain conditions) is deferred to Part II of this study.

## 1. Introduction

A so-called 'extreme material' possesses mechanical properties (typically an extreme degree of anisotropy) designed in such a way as to keep the material in a state close to an instability threshold (for instance failure of ellipticity), so that ultimate mechanical effects are displayed, which can be exploited in different technologies, for instance, in mechanical wave guiding, stress wave shielding, and invisibility cloaking. An example of a material with extreme properties is the pinscreen, a toy (invented by W. Fleming, Fig. 1) made up of a perforated plate with each hole filled with a movable pin. The amazing behavior of the pinscreen, which is the key of the commercial success of the toy, is related to the possibility of experiencing 'by hand' the strange properties of an extreme material, in which the *load does not diffuse*, but remains confined to the load indenter. Technically, the pinscreen is a material working at the threshold of loss of ellipticity.

---

[1] Corresponding author. Email: bigoni@ing.unitn.it

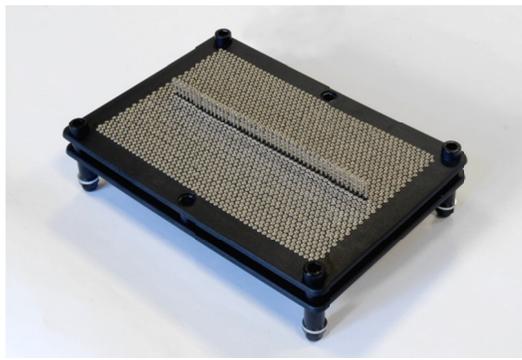

**Fig.1**: Quasi-static behavior of a material (so-called 'pinscreen', invented by W. Fleming) working at the boundary of ellipticity loss: a line loading does not diffuse, showing a strong discontinuity.

Theoretically, stress channelling was pioneered by Everstine and Pipkin (1971), Pipkin (1984), and Spencer (1984), who demonstrated that stress does not diffuse in materials with extreme orthotropic stiffness ratios, and was found to occur in masonry models by Bigoni and Noselli (2010a;b). In the limit when the stiffness ratio between different material directions tends to zero, the equations governing equilibrium reach the elliptic boundary and the stress percolates through null-thickness deformation bands. In this situation, the material microstructure sets the percolation thickness and becomes a dominant factor.

The purpose of the present article is to analyze stress channelling and strain localization, in extremely anisotropic elastic materials governed by couple-stress elasticity. The theory of couple-stress elasticity, also known as Cosserat theory with constrained rotations, is the simplest gradient theory in which couple-stresses make their appearance. In particular, couple-stress theory assumes an augmented form of the Euler-Cauchy principle with a non-vanishing couple traction, and a strain-energy density that depends upon both the strain and the gradient of rotation. Such assumptions are appropriate for materials with granular and layered structure, where the interaction between adjacent elements may introduce internal moments. In this way, characteristic material lengths appear that can be related with the material microstructure (Yang and Lakes, 1982; Zhang and Sharma, 2005; Maranganti and Sharma, 2007). The presence of these characteristic lengths implies that the couple-stress theory encompasses the analytical possibility of size effects which are absent in the classical theory. Therefore, the couple-stress theory and, more in general gradient and polar type theories, have been often advocated (Triantafyllidis and Aifantis, 1986; Mühlhaus and Vardoulakis, 1987; Gao et al. 1999; Huang et al. 2000; Radi, 2003; Radi and Gei, 2004; Dal Corso and Willis, 2011; Fleck et al., 2014) as possible remedies to constitutive formulations where ellipticity is lost in the classical sense and so strain localization into vanishing narrow bands is predicted to occur (with the related well-known mesh sensitivity in finite element formulations, Needleman and Tvergaard, 1984; Petryk, 1997)

Despite the large number of articles devoted to the analysis of gradient effects in localization problems, the condition of failure of ellipticity has neither been obtained for constrained Cosserat elasticity, nor has it been linked to the possible emergence of localized solutions of boundary value problems or to the condition of planar wave propagation. More in general, with the exception of the 'Kirchhoff-like' uniqueness statement (which was related to loss of positive definiteness of the strain energy by Mindlin and Tiersten, 1962; see also Grentzelou and Georgiadis, 2005), material instability thresholds –for instance, strong ellipticity – have never been analyzed for constrained Cosserat elasticity. However, mention should be made to related works in the context of micropolar (unconstrained Cosserat) theory, where the ellipticity and strong ellipticity conditions have been derived, the latter providing a criterion for the existence of acceleration waves (Eremeyev, 2005; Altenbach et al., 2010; Eremeyev et al., 2013).

With reference to an anisotropic theory of couple-stress elasticity, the purpose of the Part I of the present article is: (i.) to *introduce* the notion of strong ellipticity and (ii.) to motivate this through an extension of the van Hove theorem (van Hove, 1947; Hayes, 1966; Bigoni and Zaccaria, 1992) as the condition of uniqueness for the homogeneous boundary value problem with prescribed kinematical conditions; (iii.) to obtain the condition for propagation of planar waves; (iv.) to derive the condition of ellipticity related to the structure of the differential operator of the governing field equations; (v.) to establish a hierarchy between the mentioned material stability criteria; (vi.) to obtain the condition holding at a discontinuity plane and to show how these conditions are related to the failure of ellipticity; (vii.) to analyze the particular case of antiplane elasticity where a classification into Elliptic-Imaginary, Elliptic-Complex, Hyperbolic, and Parabolic regimes is introduced; viii.) to provide examples of extreme materials where waves can propagate but ellipticity is lost.

In order to complete the picture of material instabilities in Cosserat anisotropic elasticity and to relate this to the design of extreme materials, Part II of this article is devoted to the derivation of infinite-body Green's functions that will be shown to be related to the condition of wave propagation. The pertinent Green's functions will then be used to show, in proximity of the border of ellipticity loss, the emergence of folding and faulting of a continuum in simple and cross geometries.

## 2. Fundamentals of constrained couple-stress anisotropic elasticity

In this Section, the equations governing the linearized elastic mechanical response are introduced for anisotropic couple-stress solids. Detailed presentations of the couple-stress theory have been given by Toupin (1962), and Mindlin and Tiersten (1962) (see also Muki and Sternberg, 1965; Gourgiotis and Piccolroaz, 2014). An extension of the couple-stress theory to finite deformations has been recently given by Srinivasa and Reddy (2013).

Couple-stress elasticity assumes that: (i.) each material particle has three degrees of freedom, (ii.) an augmented form of the Euler-Cauchy principle holds in which a non-vanishing couple traction prevails, and (iii.) the strain-energy density depends upon both strain and the rotation gradient.

In the absence of inertia and rotary inertia effects, the balance laws for the linear and angular momentum read

$$\int_{\partial B} T_q^{(n)} dS + \int_B X_q \, dB = 0, \tag{1}$$

$$\int_{\partial B} \left( e_{qpk} x_p T_k^{(n)} + M_q^{(n)} \right) dS + \int_B \left( e_{qpk} x_p X_k + Y_q \right) dB = 0, \tag{2}$$

where $B$ is the region (open set) occupied by the body with bounding surface $\partial B$, possessing a unique outward normal $n_q$. Note that a Cartesian rectangular coordinate system is employed along with indicial notation and summation convention on repeated indices. In addition, $e_{qpk}$ is the Levi-Civita alternating symbol, $T_q^{(n)}$ is the surface force per unit area, $M_q^{(n)}$ the surface moment per unit area, $X_q$ the body force per unit volume, $Y_q$ the body moment per unit volume, and $x_p$ designates the components of the position vector.

The *stress* and *couple-stress* tensors are introduced by considering the equilibrium of the elementary material tetrahedron and enforcing Eqs. (1) and (2), respectively (Malvern, 1969)

$$T_q^{(n)} = \sigma_{pq} n_p, \quad M_q^{(n)} = \mu_{pq} n_p, \tag{3}$$

where $\sigma_{pq}$ and $\mu_{pq}$ are the components of the stress tensor and couple-stress tensor (both being asymmetric), respectively. In addition, by assuming that the interaction across any internal surface consists of equal, opposite, and collinear forces plus equal and opposite couples, it can be readily shown that the Newton's law of action and reaction holds, namely, $\mathbf{T}^{(n)} = -\mathbf{T}^{(-n)}$ and $\mathbf{M}^{(n)} = -\mathbf{M}^{(-n)}$ (Truesdell and Toupin, 1960; Malvern, 1969). Using Eqs. (2) and applying the divergence theorem, the following force and moment equations of equilibrium are obtained

$$\sigma_{pq,p} + X_q = 0, \tag{4}$$

$$e_{qpk}\sigma_{pk} + \mu_{pq,p} + Y_q = 0. \tag{5}$$

Further, the stress tensor $\sigma_{pq}$ is resolved into its symmetric and anti-symmetric components as

$$\sigma_{pq} = \tau_{pq} + \alpha_{pq}, \tag{6}$$

with $\tau_{pq} = \tau_{qp}$ and $\alpha_{pq} = -\alpha_{qp}$, whereas it is advantageous to decompose $\mu_{pq}$ into its deviatoric $m_{pq}$ and spherical $(\mu_{kk}/3)\delta_{pq}$ parts, respectively

$$\mu_{pq} = m_{pq} + \frac{1}{3}\delta_{pq}\mu_{kk}, \tag{7}$$

where $\delta_{pq}$ is the Kronecker delta. From the moment equation of equilibrium (5), one may further obtain the anti-symmetric part of the stress tensor as

$$\alpha_{pq} = -\frac{1}{2}e_{pqk}\left(\mu_{rk,r} + Y_k\right), \qquad (8)$$

so that the stress tensor becomes symmetric in the absence of body moments and for a vanishing divergence of couple-stresses. Moreover, a combination of Eqs. (5)-(8) yields a *single* equation of equilibrium which involves only the symmetric part of the stress tensor and the deviatoric part of the couple-stress tensor as

$$\tau_{pq,p} - \frac{1}{2}e_{pqk}\left(m_{rk,rp} + Y_{k,p}\right) + X_q = 0. \qquad (9)$$

For the kinematical description of the continuum, the following primary kinematical fields are defined within the framework of a geometrically linear theory

$$\varepsilon_{pq} = \frac{1}{2}\left(u_{q,p} + u_{p,q}\right), \quad \omega_q = \frac{1}{2}e_{qpk}u_{k,p}, \quad \kappa_{pq} = \omega_{q,p}, \qquad (10)$$

where $\varepsilon_{pq}$ is the strain tensor, $\omega_q$ the rotation vector, and $\kappa_{pq}$ is the curvature tensor (i.e. the transpose of the gradient of rotation $\boldsymbol{\kappa} = \nabla\boldsymbol{\omega}^T$ – note that the notation of Mindlin and Tiersten (1962) is adopted in the present work) which by definition is traceless, $\kappa_{pp} = 0$. Accordingly, the compatibility equations for the above kinematical fields are (Naghdi, 1965)

$$e_{pqm}e_{mnk}\varepsilon_{pn,k} + e_{qnm}\kappa_{mn} = 0, \quad e_{qkm}e_{npm}\kappa_{qn,k} = 0, \qquad (11)$$

where the elimination of $\kappa_{pq}$ from Eqs. (11) leads to the usual Saint Venant's compatibility equations for the strain tensor components.

Regarding the boundary conditions in the constrained couple-stress theory, it is noted that the normal component of the rotation on the boundary $\partial B$ is *fully* specified by the distribution of tangential displacements. This implies that the *tangential* part of the rotation vector, $\left(\delta_{pq} - n_p n_q\right)\omega_p$, has to be prescribed independently on the surface when defining kinematical boundary conditions. Therefore, the number of the kinematical boundary conditions that can be specified on the boundary are *five*: the three components of the displacement $u_q$ and the two tangential components of the rotation $\omega_q$. Accordingly, the traction boundary conditions at any point on a *smooth* boundary, consist of the following *three reduced* force-tractions and *two tangential* couple-tractions (Mindlin and Tiersten, 1962; Koiter, 1964)

$$P_q^{(n)} = \sigma_{pq} n_p - \frac{1}{2} e_{qpk} n_p m_{(nn),k}, \quad R_q^{(n)} = m_{pq} n_p - m_{(nn)} n_q, \tag{12}$$

where $m_{(nn)} = n_p n_q m_{pq}$ is the normal component of the deviatoric couple-stress tensor $m_{pq}$. It is worth noting that the spherical part of the couple-stress tensor $\mu_{kk}$ does not appear either in the field equations (9) or in the boundary conditions (12). It follows that this quantity remains indeterminate in the constrained Cosserat theory. Following Muki and Sternberg (1965), and without loss in generality, we shall henceforth adopt the normalization: $\mu_{kk} = 0$.

For linear constitutive behavior, the strain-energy density assumes the following general quadratic form for *centrosymmetric* materials

$$W = \frac{1}{2}\mathbb{C}_{pqmn}\varepsilon_{pq}\varepsilon_{mn} + \frac{1}{2}\mathbb{B}_{pqmn}\kappa_{pq}\kappa_{mn}, \tag{13}$$

where $\mathbb{C}_{pqmn}$ and $\mathbb{B}_{pqmn}$ are the elasticity tensors with the following symmetries (Mindlin and Tiersten, 1962)

$$\mathbb{C}_{pqmn} = \mathbb{C}_{mnpq} = \mathbb{C}_{qpnm}, \quad \mathbb{B}_{pqmn} = \mathbb{B}_{mnpq}, \quad \mathbb{B}_{pqmm} = \mathbb{B}_{mmpq} = 0. \tag{14}$$

The last equality follows directly from the fact that the curvature tensor is deviatoric ($\kappa_{pp} = 0$). Accordingly, tensor $\mathbb{C}$ is defined through 21 independent components, whereas $\mathbb{B}$ has 36 independent components. The corresponding constitutive equations are

$$\tau_{pq} = \frac{\partial W}{\partial \varepsilon_{pq}} = \mathbb{C}_{pqmn}\varepsilon_{mn}, \quad m_{pq} = \frac{\partial W}{\partial \kappa_{pq}} = \mathbb{B}_{pqmn}\kappa_{mn}. \tag{15}$$

It is remarked that tensor $\mathbb{C}$ defines a Cauchy (or 'classical') elastic behavior, which is recovered when tensor $\mathbb{B}$, defining a 'purely Cosserat behavior', vanishes.

The necessary and sufficient conditions for the strain energy density $W$ in Eq. (13) to be positive definite (PD) are

$$\mathbb{C}_{pqmn}\varepsilon_{pq}\varepsilon_{mn} > 0 \quad \forall \varepsilon_{pq} \in \text{Sym} \setminus \{0\}, \quad \mathbb{B}_{pqmn}\kappa_{pq}\kappa_{mn} > 0 \quad \forall \kappa_{pq} \in \text{Dev} \setminus \{0\}, \tag{16}$$

where Sym denotes the set of all symmetric tensors, Dev is the set of all deviatoric tensors, and 0 denotes the null element, which is excluded from the definition of positive definiteness. The condition of (PD) is sufficient for unconditional *stability* and *uniqueness* of the solution of the mixed boundary value problem (Mindlin and Tiersten, 1962; Grentzelou and Georgiadis, 2005). However, in classical elasticity lack of positive definiteness of the strain energy density is a subject of increasing attention (Lakes and Drugan, 2002; Kochmann and Drugan, 2009) as it can model situations in which prestressed solids release energy (and thus apparently violate thermodynamic requirements).

Finally, incorporating the constitutive equations (15) into Eqs. (9) and using the geometric relations in (10) yields the equations of equilibrium in terms of the displacement components (the counterpart of the Navier-Cauchy equations of the classical theory)

$$\mathbb{C}_{pqmn}u_{n,mp} - \frac{1}{4}e_{pqk}e_{smn}\mathbb{B}_{rkts}u_{n,mtrp} - \frac{1}{2}e_{pqk}Y_{k,p} + X_q = 0. \tag{17}$$

## 3. Strong ellipticity and van Hove uniqueness theorem

The following definitions of strong ellipticity (SE) for the elasticity tensors $\mathbb{C}$ and $\mathbb{B}$ are introduced, respectively

$$(\mathbf{q}\otimes\mathbf{n})\cdot\mathbb{C}[\mathbf{q}\otimes\mathbf{n}] > 0, \quad (\mathbf{q}\otimes\mathbf{n})\cdot\mathbb{B}[\mathbf{q}\otimes\mathbf{n}] > 0, \tag{18}$$

to be satisfied for every unit vector **n** and non-null vector **q**. In view of the properties of the elasticity tensors defined by Eqs. (14), the conditions of (SE) for the classical and couple-stress elasticity tensors can be equivalently written as

$$\text{Sym}(\mathbf{q} \otimes \mathbf{n}) \cdot \mathbb{C}\left[\text{Sym}(\mathbf{q} \otimes \mathbf{n})\right] > 0, \quad \text{Dev}(\mathbf{q} \otimes \mathbf{n}) \cdot \mathbb{B}\left[\text{Dev}(\mathbf{q} \otimes \mathbf{n})\right] > 0, \tag{19}$$

respectively, which, according to Eqs. (16), allow for the conclusion that if the elasticity tensors are positive definite (PD) they are also strongly elliptic

$$(\text{PD})^{\mathbb{C}} \Rightarrow (\text{SE})^{\mathbb{C}} \quad \text{and} \quad (\text{PD})^{\mathbb{B}} \Rightarrow (\text{SE})^{\mathbb{B}}. \tag{20}$$

A constrained Cosserat material is defined to be *strongly elliptic* if both inequalities in Eqs. (19) are simultaneously satisfied. Note that by replacing '>' with '≥', the conditions (19) define the 'semi-strong ellipticity' (SSE) conditions for $\mathbb{C}$ and $\mathbb{B}$.

It will now be proven that: *the (SE) conditions (19) are sufficient for uniqueness for a problem with prescribed displacement and rotation on the whole boundary (kinematical boundary conditions) in a homogeneous constrained Cosserat solid (called 'van Hove conditions')*. This statement represents the extension of the van Hove's theorem (van Hove, 1947) to the context of the constrained Cosserat theory.

Following Hayes (1966) and Gurtin, (1972), and invoking superposition, if two solutions for displacements and rotations are possible in the van Hove conditions, the solution produced by the difference fields would correspond to homogeneous conditions at the boundary (for both displacement and rotations) and would make the strain energy $U(\boldsymbol{\varepsilon}, \boldsymbol{\kappa})$ to vanish

$$2U(\boldsymbol{\varepsilon}, \boldsymbol{\kappa}) = \int_B \boldsymbol{\varepsilon} \cdot \mathbb{C}[\boldsymbol{\varepsilon}] dB + \int_B \boldsymbol{\kappa} \cdot \mathbb{B}[\boldsymbol{\kappa}] dB$$
$$= \int_B \nabla \mathbf{u} \cdot \mathbb{C}[\nabla \mathbf{u}] dB + \int_B \nabla \boldsymbol{\omega}^T \cdot \mathbb{B}[\nabla \boldsymbol{\omega}^T] dB = 0. \tag{21}$$

Therefore, to prove uniqueness it suffices to show that

$$\mathbf{X} \equiv \mathbf{0}, \ \mathbf{Y} \equiv \mathbf{0} \ \text{on } B \quad \text{and} \quad \mathbf{u} = \mathbf{0}, \ (\mathbf{I} - \mathbf{n} \otimes \mathbf{n})\boldsymbol{\omega} = \mathbf{0} \ \text{on } \partial B, \tag{22}$$

imply that $\mathbf{u} = \mathbf{0}$ (and consequently $\boldsymbol{\omega} = \mathbf{0}$) on $B$. Note that since the surface displacements are zero, the normal component of the surface rotation vector is also zero and, thus, the condition $(\mathbf{I} - \mathbf{n} \otimes \mathbf{n})\boldsymbol{\omega} = \mathbf{0}$ is equivalent to $\boldsymbol{\omega} = \mathbf{0}$ on $\partial B$.

The definitions of $\mathbf{u}$ and $\boldsymbol{\omega}$ can be extended from the closed set $\bar{B}$ to the whole Euclidean space $\Re$ simply by defining: $\mathbf{u} = \mathbf{0}$ and $\boldsymbol{\omega} = \mathbf{0}$ on $\Re \setminus \bar{B}$. The resulting fields are piecewise continuous on $\Re$, whereas the gradient of the displacement $\nabla \mathbf{u}$ and the gradient of the rotation $\nabla \boldsymbol{\omega}$ may be discontinuous on $\partial B$. In particular, in view of Eq. (10)$_2$ and bearing in mind that the surface rotation is zero, it can be readily shown that only the part $\nabla \mathbf{u} \cdot (\mathbf{n} \otimes \mathbf{n})$ of the displacement gradient can be discontinuous on $\partial B$. The vectors $\mathbf{u}$, $\boldsymbol{\omega}$, and their gradients possess the following three-dimensional Fourier representations

$$\tilde{\mathbf{u}}(\mathbf{k}) = \frac{1}{(2\pi)^3} \int_\Re \mathbf{u}(\mathbf{x}) e^{i\mathbf{k}\cdot\mathbf{x}} d^3\mathbf{x}, \quad \mathbf{U}(\mathbf{k}) = \frac{1}{(2\pi)^3} \int_\Re \nabla \mathbf{u}(\mathbf{x}) e^{i\mathbf{k}\cdot\mathbf{x}} d^3\mathbf{x}, \quad (23)$$

$$\tilde{\boldsymbol{\omega}}(\mathbf{k}) = \frac{1}{(2\pi)^3} \int_\Re \boldsymbol{\omega}(\mathbf{x}) e^{i\mathbf{k}\cdot\mathbf{x}} d^3\mathbf{x}, \quad \boldsymbol{\Omega}(\mathbf{k}) = \frac{1}{(2\pi)^3} \int_\Re \nabla \boldsymbol{\omega}(\mathbf{x}) e^{i\mathbf{k}\cdot\mathbf{x}} d^3\mathbf{x}, \quad (24)$$

where $d^3\mathbf{x} = dx_1 dx_2 dx_3$, $\mathbf{k}$ is the Fourier vector in the Fourier space, and $i = (-1)^{1/2}$. Utilizing the kinematical boundary conditions in Eq. (22) and the divergence theorem, it can be readily shown that

$$\mathbf{U} = -i\tilde{\mathbf{u}} \otimes \mathbf{k}, \quad \boldsymbol{\Omega} = -i\tilde{\boldsymbol{\omega}} \otimes \mathbf{k}. \quad (25)$$

Moreover, since the body is homogeneous, the generalized Parseval theorem yields (Davies, 2002)

$$\int_\Re \nabla \mathbf{u} \cdot \mathbb{C}[\nabla \mathbf{u}] d^3\mathbf{x} = \mathbb{C}_{pqmn} \int_\Re u_{p,q} u_{m,n} d^3\mathbf{x} = \mathbb{C}_{pqmn} \int_\Re U_{pq} \bar{U}_{mn} d^3\mathbf{k} = \int_\Re \mathbf{U} \cdot \mathbb{C}[\bar{\mathbf{U}}] d^3\mathbf{k}, \quad (26)$$

$$\int_\Re \nabla \boldsymbol{\omega}^\mathrm{T} \cdot \mathbb{B}[\nabla \boldsymbol{\omega}^\mathrm{T}] d^3\mathbf{x} = \mathbb{B}_{pqmn} \int_\Re \omega_{q,p} \omega_{n,m} d^3\mathbf{x} = \mathbb{B}_{pqmn} \int_\Re \Omega_{qp} \bar{\Omega}_{nm} d^3\mathbf{k} = \int_\Re \boldsymbol{\Omega}^\mathrm{T} \cdot \mathbb{B}[\bar{\boldsymbol{\Omega}}^\mathrm{T}] d^3\mathbf{k}, (27)$$

with $\bar{\mathbf{U}}$ and $\bar{\boldsymbol{\Omega}}$ being the complex conjugates of $\mathbf{U}$ and $\boldsymbol{\Omega}$, respectively. Note that Eq. (25) implies

$$\mathbf{U}_R = \tilde{\mathbf{u}}_I \otimes \mathbf{k}, \quad \mathbf{U}_I = -\tilde{\mathbf{u}}_R \otimes \mathbf{k} \quad \text{and} \quad \boldsymbol{\Omega}_R = \tilde{\boldsymbol{\omega}}_I \otimes \mathbf{k}, \quad \boldsymbol{\Omega}_I = -\tilde{\boldsymbol{\omega}}_R \otimes \mathbf{k}, \tag{28}$$

where the indices $R$ and $I$ stand for the real and imaginary parts, respectively. A substitution of Eqs. (28) into Eqs. (26) and (27) yields

$$\int_B \nabla \mathbf{u} \cdot \mathbb{C}[\nabla \mathbf{u}] d^3 \mathbf{x} = \int_{\mathfrak{R}} (\tilde{\mathbf{u}}_R \otimes \mathbf{k}) \cdot \mathbb{C}[\tilde{\mathbf{u}}_R \otimes \mathbf{k}] d^3 \mathbf{k} + \int_{\mathfrak{R}} (\tilde{\mathbf{u}}_I \otimes \mathbf{k}) \cdot \mathbb{C}[\tilde{\mathbf{u}}_I \otimes \mathbf{k}] d^3 \mathbf{k}, \tag{29}$$

$$\int_B \nabla \boldsymbol{\omega}^{\mathrm{T}} \cdot \mathbb{B}[\nabla \boldsymbol{\omega}^{\mathrm{T}}] d^3 \mathbf{x} = \int_{\mathfrak{R}} (\mathbf{k} \otimes \tilde{\boldsymbol{\omega}}_R) \cdot \mathbb{B}[\mathbf{k} \otimes \tilde{\boldsymbol{\omega}}_R] d^3 \mathbf{k} + \int_{\mathfrak{R}} (\mathbf{k} \otimes \tilde{\boldsymbol{\omega}}_I) \cdot \mathbb{B}[\mathbf{k} \otimes \tilde{\boldsymbol{\omega}}_I] d^3 \mathbf{k}. \tag{30}$$

Assume now that the fields $\mathbf{u}$ and $\boldsymbol{\omega}$ do not vanish identically in $B$, then by the uniqueness of the Fourier transforms, $\tilde{\mathbf{u}}$ and $\tilde{\boldsymbol{\omega}}$ cannot vanish identically on $\mathfrak{R}$. Hence, in view of the Eqs. (29) and (30), if the elasticity tensors $\mathbb{C}$ and $\mathbb{B}$ are strongly elliptic then $U(\boldsymbol{\varepsilon}, \boldsymbol{\kappa}) > 0$, which contradicts the original assumption in Eq. (21). Therefore, when (SE) holds for a *homogeneous* Cosserat material, uniqueness of the kinematical b.v.p. follows. The importance of the van Hove theorem lies in the fact that it provides conditions to reach the limit of stress channelling (or shear banding, in the context of elastoplasticity), without encountering prior bifurcations (see Ryzhak, 1987; Bigoni and Zaccaria, 1992).

In the next Section, the (SE) conditions (19) will be proven to be *sufficient but not necessary conditions* to ensure propagation of plane waves in a couple-stress medium. This is in marked contrast with the classical elasticity case where the (SE) of the Cauchy elasticity tensor $\mathbb{C}$ is a *necessary and sufficient* condition for the positive definiteness of the acoustic tensor which, in turn, implies that all waves can travel with real and positive propagation speed (Gurtin, 1972).

## 4. Wave propagation and acoustic tensor

The propagation of plane waves in couple-stress elasticity is now considered. In particular, a plane harmonic wave is defined through a vector field of the form

$$\mathbf{u} = \mathrm{Re}\left[\mathbf{d} \, e^{-i(k\mathbf{x} \cdot \mathbf{n} - \omega t)}\right], \tag{31}$$

where $t$ denotes time, $\mathbf{d}$ the wave amplitude vector, $\mathbf{n}$ the unit propagation vector, and $k$ the wavenumber (in general complex). Moreover, the vector $\mathbf{x}$ denotes the position vector, $\omega = kV$ is the angular frequency assumed to be always real, and $V$ is the phase velocity. The equations of motion in couple-stress elasticity follow from Eqs. (17), augmenting the latter with the standard inertia term

$$\mathbb{C}_{pqmn}u_{n,mp} - \frac{1}{4}e_{pqk}e_{smn}\mathbb{B}_{rkts}u_{n,mtrp} = \rho \ddot{u}_q, \qquad (32)$$

where $\rho > 0$ is the constant mass density, a superposed dot denotes time differentiation, and null body forces and moments are assumed. Note that for the purposes of the present analysis, the micro-rotational inertia effects are neglected.

A substitution of Eq. (31) into the equations of motion leads to the propagation condition

$$\left(\mathbf{A}(k,\mathbf{n}) - \rho\omega^2\mathbf{I}\right)\mathbf{d} = \mathbf{0}, \qquad (33)$$

where the *Cosserat acoustic tensor* $\mathbf{A}$ can be decomposed into a classical part $\mathbf{A}^{(\mathbb{C})}$ and an additional couple-stress part $\mathbf{A}^{(\mathbb{B})}$ in the following way (Toupin, 1962)

$$\mathbf{A}(k,\mathbf{n}) = k^2\mathbf{A}^{(\mathbb{C})}(\mathbf{n}) + k^4\mathbf{A}^{(\mathbb{B})}(\mathbf{n}), \qquad (34)$$

with

$$A_{qn}^{(\mathbb{C})}(\mathbf{n}) = \mathbb{C}_{pqmn}n_p n_m \quad \text{and} \quad A_{qn}^{(\mathbb{B})}(\mathbf{n}) = \frac{1}{4}e_{pqk}e_{smn}n_m n_p n_t n_r \mathbb{B}_{rkts}. \qquad (35)$$

Note that the symmetries of the elasticity tensors $\mathbb{C}$ and $\mathbb{B}$ imply also that $\mathbf{A}^{(\mathbb{C})}$ and $\mathbf{A}^{(\mathbb{B})}$ are symmetric second-order tensors. The symmetry of $\mathbf{A}^{(\mathbb{B})}$ follows from the fact that $e_{lqn}A_{qn}^{(\mathbb{B})} = 0$. Therefore, the acoustic tensor in couple-stress theory is *symmetric*, $\mathbf{A} = \mathbf{A}^{\mathrm{T}}$. Moreover, it is remarked that the components of the acoustic tensor are *non-homogeneous* polynomials of fourth-degree with respect to the wave number $k$. Hence, contrary to the classical elasticity case, the frequency and the phase velocity depend on the wave number, which implies, in general, that waves are dispersive in the context of the couple-stress theory.

The condition (33) dictates that the amplitude $\mathbf{d}$ must be an eigenvector of the acoustic tensor $\mathbf{A}$ while its eigenvalues $\omega^2$ (to within a multiplicative constant $\rho$) are always real and the respective eigenspaces orthogonal. A non-trivial solution exists for the eigenvalue problem (33) when the determinant of the coefficients $d_q$ vanishes,

$$\det\left(\mathbf{A} - \rho\omega^2 \mathbf{I}\right) = 0. \tag{36}$$

An important property of the acoustic tensor is that its couple-stress part $\mathbf{A}^{(\mathbb{B})}$ is singular, so that the propagation vector $\mathbf{n}$ is always an eigenvector of $\mathbf{A}^{(\mathbb{B})}$ associated to a *null* eigenvalue

$$\det \mathbf{A}^{(\mathbb{B})} = 0, \quad \mathbf{A}^{(\mathbb{B})} \mathbf{n} = \mathbf{0}, \tag{37}$$

and consequently,

$$\mathbf{A}\mathbf{n} = k^2 \mathbf{A}^{(\mathbb{C})} \mathbf{n}, \tag{38}$$

which implies that for every couple-stress anisotropy $\mathbb{B}_{rkts}$ at least one direction of propagation $\mathbf{n}$ exists such that the wave characteristics are governed only by the Cauchy-elastic part of the constitutive equations. This direction coincides with the direction of propagation of purely longitudinal $P$-waves in a *classical anisotropic* medium. In fact, as it was shown by Truesdell (1966) (see also Kolodner, 1966) in the context of the classical theory, there always exist at least three distinct directions along which longitudinal waves can propagate. Therefore: *regardless to the degree of anisotropy, a longitudinal $P$-wave always exists propagating in a Cosserat elastic material without dispersion and without displaying Cosserat effects (provided this wave can propagate in the underlying Cauchy material).*

A necessary and sufficient condition for plane waves to propagate with positive speed and for all real wavenumbers $k$ is that *the acoustic tensor is positive definite*, which is now defined as the (WP) condition

$$\mathbf{p} \cdot \mathbf{A}(k, \mathbf{n}) \mathbf{p} > 0 \Leftrightarrow k^2 \left( \mathbf{p} \cdot \mathbf{A}^{(\mathbb{C})} \mathbf{p} + k^2 \mathbf{p} \cdot \mathbf{A}^{(\mathbb{B})} \mathbf{p} \right) > 0, \tag{39}$$

for every unit vector $\mathbf{n}$ and non-null vector $\mathbf{p}$. Thus, if the wave propagation (WP) condition holds, the squared speeds (i.e. the eigenvalues of the acoustic tensor) corresponding to each real acoustical axis are positive. Accordingly, three linearly independent plane waves always exist for a given direction of propagation $\mathbf{n}$ and wavenumber $k$. It is apparent from Eq. (39), that for small wavenumbers $k \to 0$ (low frequencies) the classical part $\mathbf{A}^{(C)}$ dominates the behavior of the acoustic tensor, whereas for large wavenumbers $k \to \infty$ (high frequencies) the behavior of the acoustic tensor is determined by its couple-stress part $\mathbf{A}^{(B)}$. Therefore, by taking into account that Eq. (39) must hold for all real non-zero wavenumbers, the (WP) condition in couple-stress elasticity is equivalent to the following pair of inequalities

$$\mathbf{p} \cdot \mathbf{A}^{(C)} \mathbf{p} \geq 0, \quad \mathbf{p} \cdot \mathbf{A}^{(B)} \mathbf{p} \geq 0, \quad \mathbf{p} \cdot \mathbf{A} \mathbf{p} \neq 0, \quad \forall \mathbf{p} \neq \mathbf{0}, \tag{40}$$

so that both '$=0$' cannot simultaneously apply in Eqs. (40)$_1$ and (40)$_2$. In fact, the above conditions imply that $\mathbf{p}$ *cannot* be an eigenvector corresponding to a *null* eigenvalue of both the classical part $\mathbf{A}^{(C)}$ and the couple-stress part $\mathbf{A}^{(B)}$ of the acoustic tensor. An immediate consequence of this statement is that when $\mathbf{p}$ is parallel to the propagation vector $\mathbf{n}$, Eqs. (40) reduce to

$$\mathbf{n} \cdot \mathbf{A}^{(C)} \mathbf{n} > 0, \tag{41}$$

which excludes the possibility that $\mathbf{n}$ be an eigenvector of $\mathbf{A}^{(C)}$ corresponding to a null eigenvalue.

On the other hand, in view of the inequality (18)$_1$, it can be readily shown that the (SE) of the classical Cauchy elasticity tensor is equivalent to the condition that $\mathbf{A}^{(C)}$ be positive definite

$$(SE)^{\mathbb{C}} \Leftrightarrow \mathbf{p} \cdot \mathbf{A}^{(C)} \mathbf{p} > 0, \quad \forall \mathbf{p} \neq \mathbf{0}. \tag{42}$$

Further, by setting $\mathbf{q} = \mathbf{n} \times \mathbf{p}$ and taking into account Eq. (35)$_2$, the (SE) condition of the couple-stress elasticity tensor, Eq. (18)$_2$, implies

$$(\text{SE})^{\mathbb{B}} \Rightarrow \mathbf{p} \cdot \mathbf{A}^{(\mathbb{B})} \mathbf{p} > 0, \quad \forall \mathbf{p} : \mathbf{n} \times \mathbf{p} \neq \mathbf{0}, \tag{43}$$

so that, since $\mathbf{A}^{(\mathbb{B})}$ is symmetric and possesses always one null eigenvalue associated with the vector $\mathbf{n}$ (say, $\lambda_1 \equiv 0$), the condition $(\text{SE})^{\mathbb{B}}$ is equivalent to the fact that the remaining two eigenvalues must be strictly positive (i.e. $\lambda_2 > 0$ and $\lambda_3 > 0$). Replacing the strict inequalities in Eqs. (42) and (43) with "$\geq$", the above conditions become the semi-strong ellipticity (SSE) conditions, holding for every $\mathbf{p}$

$$(\text{SSE})^{\mathbb{C}} \Leftrightarrow \mathbf{p} \cdot \mathbf{A}^{(\mathbb{C})} \mathbf{p} \geq 0, \qquad (\text{SSE})^{\mathbb{B}} \Leftrightarrow \mathbf{p} \cdot \mathbf{A}^{(\mathbb{B})} \mathbf{p} \geq 0. \tag{44}$$

Note that the inequality in (44)$_2$ has been extended now to include the case where $\mathbf{p}$ is parallel to the propagation vector $\mathbf{n}$.

The above results show that the (SE) of the elasticity tensors is a *sufficient* condition to guarantee the propagation of plane waves (WP). The *necessary and sufficient* conditions for (WP) are expressed through conditions (40), which are equivalent to the positive definiteness of the acoustic tensor $\mathbf{A}$. These conditions are weaker than those implied by (SE). It will be shown in Part II that the (WP) plays a major role for the derivation of the infinite body Green's function.

*4.1 Wave propagation in an isotropic Cosserat Material*

To illustrate the constraints imposed on the elasticities by the (WP) condition, the simple case of an isotropic couple-stress material is analyzed. In this case, the elasticity tensors defined in Eqs. (14), assume the following form (Mindlin and Tiersten, 1962; Koiter, 1964)

$$\mathbb{C}_{ijkl} = \lambda \delta_{ij}\delta_{kl} + \mu\left(\delta_{ik}\delta_{jl} + \delta_{il}\delta_{jk}\right) \quad \text{and} \quad \mathbb{B}_{ijkl} = 4\eta \delta_{ik}\delta_{jl} + 4\eta'\delta_{il}\delta_{jk} - \frac{4(\eta + \eta')}{3}\delta_{ij}\delta_{kl}, \tag{45}$$

where the moduli $\lambda$ and $\mu$ have the same meaning as the Lamé constants of the classical elasticity theory, with the dimensions of 'stress', and the moduli $\eta$ and $\eta'$ account for the couple-stress effects and are expressed in dimensions of 'force'. Upon substituting Eq. (45)

into Eq. (35), the following expressions for the classical and couple-stress parts of the acoustic tensor are obtained, respectively

$$\mathbf{A}^{(C)}(\mathbf{n}) = (\lambda + 2\mu)\mathbf{n} \otimes \mathbf{n} + \mu(\mathbf{I} - \mathbf{n} \otimes \mathbf{n}), \quad \mathbf{A}^{(B)}(\mathbf{n}) = \eta(\mathbf{I} - \mathbf{n} \otimes \mathbf{n}), \qquad (46)$$

so that the acoustic tensor can be written as

$$\mathbf{A}(k,\mathbf{n}) = k^2 \left[ (\lambda + 2\mu)\mathbf{n} \otimes \mathbf{n} + (\mu + \eta k^2)(\mathbf{I} - \mathbf{n} \otimes \mathbf{n}) \right]. \qquad (47)$$

Note that $\mu$ and $\eta$ are double eigenvalues for the tensors $\mathbf{A}^{(C)}$ and $\mathbf{A}^{(B)}$, respectively, so that every vector lying in the propagation plane characterized by $\mathbf{n}$ is an eigenvector of these tensors and, consequently, of the acoustic tensor $\mathbf{A}$. Thus, $\mathbf{A}^{(C)}$ and $\mathbf{A}^{(B)}$ are coaxial in the isotropic case. The (WP) condition (39) requires that all eigenvalues of the acoustic tensor $\mathbf{A}$ must be positive in order for the disturbance to travel with real propagation speed, a condition equivalent to

$$\lambda + 2\mu > 0, \quad \mu + \eta k^2 > 0, \quad \forall k > 0. \qquad (48)$$

The first inequality is the same as in the classical theory and is an immediate consequence of Eq. (41). In this case, the propagation vector $\mathbf{n}$ is an eigenvector of the acoustic tensor corresponding to a *non-dispersive* longitudinal *P*-wave with phase velocity $V_P = \sqrt{(\lambda + 2\mu)/\rho}$. The second inequality involves also the wave number $k$, so that either $\{\mu > 0, \eta \geq 0\}$ or $\{\mu \geq 0, \eta > 0\}$ must hold for (WP). The double eigenvalue of the acoustic tensor $\mathbf{A}$ corresponds to horizontally (SH) or vertically (SV) polarized shear waves travelling dispersively with speed and frequency

$$V_S = \sqrt{\frac{\mu + \eta k^2}{\rho}}, \quad \omega^2 = k^2 \frac{\mu + \eta k^2}{\rho}. \qquad (49)$$

Therefore, for an isotropic couple-stress material, the (WP) condition requires one of the following set of inequalities to hold

$$(\text{WP}) \Leftrightarrow \{\lambda + 2\mu > 0, \ \mu > 0, \ \eta \geq 0\} \quad \text{or} \quad \{\lambda + 2\mu > 0, \ \mu \geq 0, \ \eta > 0\}. \tag{50}$$

On the other hand, it can be readily shown that the (SE) conditions (19) require that

$$(\text{SE}) \Leftrightarrow \{\lambda + 2\mu > 0, \ \mu > 0, \ \eta > 0, \ \eta + \eta' > 0\}. \tag{51}$$

It is evident, thus, that (SE) implies the (WP) condition, but not vice-versa. Moreover, it is worth noting that for an extreme isotropic material with $\mu = 0$ and $\eta > 0$, shear type waves may still propagate. In fact, the phase velocity in Eq. (49)$_1$ becomes a linear function of the wavenumber, thus resembling the propagation of flexural harmonic waves in an Euler-Bernoulli beam (Graff, 1975)[2].

## 5. Ellipticity

The definition of ellipticity (E) is now introduced in a way appropriate for the system of partial differential equations (17). Assuming zero body forces and moments, the governing equations of equilibrium can be written as

$$\mathbf{Lu} = \mathbf{0}, \tag{52}$$

where the fourth-order differential operator $\mathbf{L}$ has the following form

$$\mathrm{L}_{qn}(\partial) \equiv \mathrm{L}_{qn}^{\mathbb{C}} + \mathrm{L}_{qn}^{\mathbb{B}} = \mathbb{C}_{pqmn} \partial_p \partial_m - \frac{1}{4} e_{pqk} e_{smn} \mathbb{B}_{rkts} \partial_m \partial_p \partial_r \partial_t, \tag{53}$$

---

[2] In the previous discussion regarding the propagation of plane waves and the associated (WP) condition, the wavenumber was taken to be a real quantity. However, as Mindlin and Tiersten (1962) pointed out, imaginary wavenumbers are also possible in the context of couple-stress elasticity. For instance, in the case of an isotropic couple-stress material with $\mu > 0$ and $\eta > 0$, the dispersion equation (49)$_2$ indicates that two wavenumbers $k$, one real and one imaginary, correspond to a given real frequency $\omega$. Accordingly, this implies that there are two shear dispersive modes: one progressive and one evanescent. The evanescent mode has a wave number cut-off at $k = \pm i\sqrt{\mu/\eta}$ corresponding to zero frequency. The existence of such evanescent modes can cause local effects near boundaries and singularities.

with $L^C_{qn}$ being the second-order (lower) part and $L^B_{qn}$ the fourth-order (higher) part of the operator. The associated *symbol* $l$ of the differential operator $\mathbf{L}$ is then defined as (Renardy and Rogers, 2004)

$$l_{qn}(i\mathbf{k}) = k^2 \mathbb{C}_{pqmn} n_p n_m + \frac{k^4}{4} e_{pqk} e_{smn} \mathbb{B}_{rkts} n_m n_p n_r n_t = k^2 A^{(C)}_{qn}(\mathbf{n}) + k^4 A^{(B)}_{qn}(\mathbf{n}), \tag{54}$$

where $\mathbf{k} = k\mathbf{n}$ is an arbitrary real vector. The (total) symbol is thus identified with the acoustic tensor, $l(i\mathbf{k}) \equiv \mathbf{A}(k, \mathbf{n})$, where its *principal part* $l^P$ is related to the highest (fourth-order) derivatives of the operator $\mathbf{L}$, with $l^P \equiv k^4 \mathbf{A}^{(B)}$. An immediate consequence of Eq. (37)$_1$ is that

$$\det l^P = 0, \tag{55}$$

showing that the principal part of the symbol is *degenerate*, so that the system of PDEs in couple-stress elasticity is not elliptic in the standard sense. To expose the degeneracy of the couple-stress operator, the divergence is applied to the governing system (52), yielding

$$\text{div}(\mathbf{L}\mathbf{u}) = \text{div}(\mathbf{L}^C \mathbf{u}) = 0, \tag{56}$$

where the higher-order operator does not appear, as $\mathbf{L}^B \mathbf{u}$ is solenoidal. Equation (56) shows that the divergence of the fourth-order couple-stress operator degenerates into a scalar third-order PDE. Moreover, applying the gradient operator to Eq. (56), a fourth-order system of PDEs is obtained

$$\text{grad}\left(\text{div}(\mathbf{L}^C \mathbf{u})\right) = \mathbf{0}. \tag{57}$$

It is remarked that Eq. (57) is equivalent to: $\tau_{qn,qn} = 0$, where $\tau_{qn}$ is the symmetric part of the stress tensor. The latter observation can also be directly derived from Eq. (9).

A *modified* couple-stress operator can now be defined by adding to the governing operator (53) the additional fourth-order operator defined in Eq. (57) as a sort of null Lagrangean

$$\tilde{L}_{qn}(\partial) = \mathbb{C}_{pqmn}\partial_p\partial_m - \frac{1}{4}e_{pqk}e_{smn}\mathbb{B}_{rkts}\partial_m\partial_p\partial_r\partial_t + a\,\mathbb{C}_{psmn}\partial_q\partial_s\partial_p\partial_m, \qquad (58)$$

where $a$ is a non-zero arbitrary constant. Note that, in view of Eq. (57), the fourth-order operators $\mathbf{L}$ and $\tilde{\mathbf{L}}$ are equivalent. The modified principal symbol is defined as

$$\tilde{l}^{\mathrm{P}} = k^4\left(\mathbf{A}^{(\mathrm{B})} - a\,\mathbf{n}\otimes\mathbf{A}^{(\mathrm{C})}\mathbf{n}\right), \qquad (59)$$

and its determinant is not trivially zero. The difficulty to classify the original degenerate operator can be now circumvented by examining the equivalent modified operator and the associated principal symbol in Eq. (59). In particular, taking into account that the principal symbol is invariant under coordinate transformations, and that $\mathbf{A}^{(\mathrm{B})}\mathbf{n} = \mathbf{0}$, the couple-stress part of the acoustic tensor admits the following spectral representation

$$\mathbf{A}^{(\mathrm{B})}(\mathbf{n}) = \lambda_2(\mathbf{n})\mathbf{t}\otimes\mathbf{t} + \lambda_3(\mathbf{n})\mathbf{s}\otimes\mathbf{s}, \qquad (60)$$

where $\lambda_2$ and $\lambda_3$ are the non-vanishing eigenvalues of $\mathbf{A}^{(\mathrm{B})}$ and $(\mathbf{n},\mathbf{t},\mathbf{s})$ are the respective eigenvectors forming an orthonormal basis. The classical part of the acoustic tensor is represented in the same orthonormal basis, so that

$$\mathbf{n}\otimes\mathbf{A}^{(\mathrm{C})}\mathbf{n} = \tau_\nu\mathbf{n}\otimes\mathbf{n} + \tau_\tau\mathbf{n}\otimes\mathbf{t} + \tau_\sigma(\mathbf{n})\mathbf{n}\otimes\mathbf{s}, \qquad (61)$$

where $\tau_\nu = \mathbf{n}\cdot\mathbf{A}^{(\mathrm{C})}\mathbf{n}$, $\tau_\tau = \mathbf{n}\cdot\mathbf{A}^{(\mathrm{C})}\mathbf{t}$, and $\tau_\sigma = \mathbf{n}\cdot\mathbf{A}^{(\mathrm{C})}\mathbf{s}$. The determinant of the principal symbol finally becomes

$$\det\tilde{l}^{\mathrm{P}} = ak^{12}\tau_\nu\lambda_2\lambda_3. \qquad (62)$$

The condition of ellipticity for a constrained Cosserat material is that $\det \tilde{\boldsymbol{l}}^{\mathrm{P}} \neq 0$, which corresponds to

$$\tau_\nu(\mathbf{n}) \neq 0, \quad \lambda_2(\mathbf{n}) \neq 0, \quad \lambda_3(\mathbf{n}) \neq 0, \quad \forall \mathbf{n}: |\mathbf{n}| = 1. \tag{63}$$

Note that two different, but equivalent ways to derive the above conditions of ellipticity are given in Appendix B.

The fact that only the conditions (63)$_2$ and (63)$_3$ refer to the Cosserat moduli, whereas condition (63)$_1$ involves the classical (Cauchy) moduli, is attributed to the degeneracy of the principal part of the symbol in couple-stress elasticity, Eq. (55). This degeneracy is directly related to the fact that in the constrained Cosserat theory the strain energy depends upon the gradient of the rotation (8 independent components) and not upon the complete gradient of strain (18 independent components). In strain gradient elasticity, where both rotation and stretch gradients are taken into account, the principal symbol is not degenerate, so that the ellipticity condition should involve only the strain gradient moduli, a problem that will be addressed elsewhere. Finally, in micropolar (unconstrained Cosserat) theory the equilibrium equations for the displacement and rotation components are of the second-order. The resulting principal symbol of the matrix operator is therefore not degenerate since the rotation vector is independent from the displacement field. The condition of ellipticity serves as a criterion for the existence of acceleration waves in a micropolar continuum (Altenbach et al., 2010; Eremeyev et al., 2013).

It is worth noting that for an isotropic couple-stress material, the (E) condition requires

$$\tau_\nu(\mathbf{n}) = \lambda + 2\mu \neq 0, \quad \lambda_2(\mathbf{n}) = \lambda_3(\mathbf{n}) = \eta \neq 0. \tag{64}$$

A comparison of the above relations with the respective (WP) conditions (50) shows that: *when $\eta = 0$ (E) is lost in the Cosserat material, but shear waves can still propagate in all directions provided that $\mu > 0$*, a circumstance in marked contrast with classical elasticity, where loss of ellipticity implies that waves cannot propagate. Indeed, for an isotropic Cauchy material with $\mu = 0$, ellipticity is lost and accordingly shear waves cannot propagate.

Finally, as it is apparent from Eqs. (42), (43) and (63), (SE) of the elasticity tensors in a Cosserat material implies (E) of the couple-stress equations. Therefore, bearing in mind the results derived in Sections 3 and 4, a hierarchy between different criteria for stability/uniqueness can be envisaged, as shown in Fig. 2.

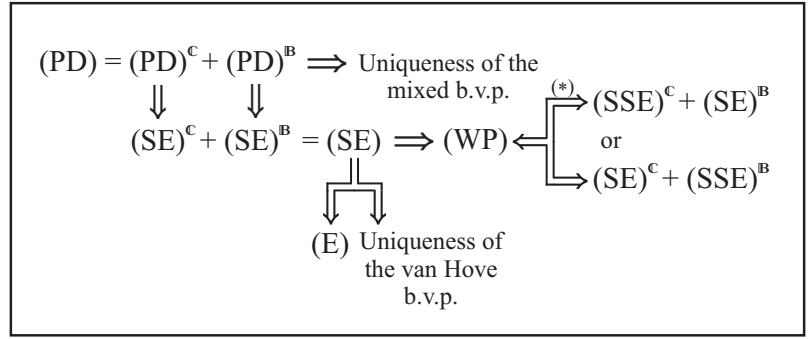

**Fig.2:** Hierarchy between the stability/uniqueness conditions: positive definiteness (PD), strong ellipticity (SE), semi-strong ellipticity (SSE), ellipticity (E), and wave propagation (WP) in a constrained Cosserat continuum. The equivalence (*) holds whenever condition (41) is verified.

## 6. Maxwell's compatibility conditions and discontinuity surfaces

Through a generalization of Thomas (1957) and Hill (1961), the compatibility conditions on discontinuity surfaces are now introduced for couple-stress elasticity. *Similarly to the 'classical' context of classical Cauchy elasticity, discontinuity surfaces will be shown to be possible at ellipticity loss.*

A smooth surface $\Sigma$ is considered with a unit normal $\mathbf{n}$, defining the common boundary of two open regions $D^+$ and $D^-$ of a solid body. Moreover, $\boldsymbol{\varphi}(\mathbf{x})$ denotes any field which is continuous in the interior of $D^+$ and $D^-$, and approaches definite limit values $\boldsymbol{\varphi}^+$ and $\boldsymbol{\varphi}^-$ as $\mathbf{x} \to \mathbf{x}_0$ with $\mathbf{x}_0 \in \Sigma$. The jump of $\boldsymbol{\varphi}$ across $\Sigma$ at $\mathbf{x}_0$ is defined as

$$[\![\boldsymbol{\varphi}]\!] = \boldsymbol{\varphi}^+\left(\mathbf{x}^+\right) - \boldsymbol{\varphi}^-\left(\mathbf{x}^-\right). \tag{65}$$

It is assumed that $\boldsymbol{\varphi}$ and all of its derivatives exist in $D^+ \cup D^-$ and have finite limiting values on $\Sigma$. For simplicity, the surface of discontinuity $\Sigma$ is assumed to be a *planar* surface.

From equilibrium considerations and imposing continuity of displacements across $\Sigma$ the following conditions are obtained

$$\left[\!\left[P_q^{(n)}\right]\!\right] = 0 , \quad \left[\!\left[R_q^{(n)}\right]\!\right] = 0 , \quad [\![u_q]\!] = 0, \tag{66}$$

where the tractions $P_q^{(n)}$ and $R_q^{(n)}$ are given by Eqs. (12), respectively. Note that Eqs. (66) do *not* include the condition of continuity of rotations, which are left for generality unprescribed across the discontinuity surface.

Denoting the normal and the surface gradients respectively as

$$D = \partial_\mathbf{n} = n_q \partial_q, \qquad D_q = \partial_q - n_q D, \tag{67}$$

and employing the Hadamard lemma (Thomas, 1957; Truesdell and Toupin, 1960; Hill, 1961), the Maxwell geometrical compatibility relations can be derived for the jumps of the first and the higher-order gradients of the displacement field in the form

$$[\![u_{n,m}]\!] = n_m g_n^{(1)}, \tag{68}$$

$$[\![u_{n,mt}]\!] = n_m n_t g_n^{(2)} + (n_t D_m + n_m D_t) g_n^{(1)}, \tag{69}$$

$$[\![u_{n,mtr}]\!] = n_r n_t n_m g_n^{(3)} + (n_r n_t D_m + n_r n_m D_t + n_t n_m D_r) g_n^{(2)}$$
$$+ (n_t D_r D_m + n_r D_m D_t + n_m D_t D_r) g_n^{(1)}, \tag{70}$$

where $\mathbf{g}^{(p)} \equiv [\![\partial_n^p \mathbf{u}]\!] = [\![\partial^p \mathbf{u}/\partial n^p]\!]$ is the discontinuity vector of the $p$-order normal derivative.

Taking now into account Eqs. (8), (12) and (15), the expressions for the continuity of tractions can be written as

$$\mathbb{C}_{pqmn} n_p [\![u_{n,m}]\!] - \frac{1}{4} e_{pqk} e_{smn} n_p \mathbb{B}_{rkts} [\![u_{n,mtr}]\!] + \frac{1}{4} e_{pqk} e_{smn} n_r n_l n_p \mathbb{B}_{rlts} [\![u_{n,mtk}]\!] = 0, \tag{71}$$

$$\frac{1}{4} e_{pqk} e_{smn} \mathbb{B}_{rkts} n_r n_p [\![u_{n,mt}]\!] = 0. \tag{72}$$

A substitution of the Maxwell compatibility conditions (68)-(70) into Eqs. (71) and (72), and subsequent use of Eqs. (35), yields a *coupled* system of differential equations relating the discontinuity vectors for a material governed by the couple-stress theory

$$\begin{cases} \left[ A^{(\mathbb{C})}_{qn} + H^{(2)}_{qn} \right] g^{(1)}_n - A^{(\mathbb{B})}_{qn} g^{(3)}_n - \tilde{H}^{(1)}_{qn} g^{(2)}_n = 0 \\ A^{(\mathbb{B})}_{qn} g^{(2)}_n + H^{(1)}_{qn} g^{(1)}_n = 0 \end{cases}, \quad (73)$$

where the *surface* gradient operators $H^{(1)}_{qj}$, $\tilde{H}^{(1)}_{qj}$, and $H^{(2)}_{qn}$ of the 1st and 2nd order are defined as

$$H^{(1)}_{qn} = \frac{1}{4} e_{pqk} e_{smn} n_r n_p \mathbb{B}_{rkts} \left( n_t D_m + n_m D_t \right), \quad (74)$$

$$\tilde{H}^{(1)}_{qn} = \frac{1}{4} e_{pqk} e_{smn} n_p n_t n_m \left( \mathbb{B}_{rkts} D_r - n_r n_l \mathbb{B}_{rlts} D_k \right) + H^{(1)}_{qn}, \quad (75)$$

$$H^{(2)}_{qn} = \frac{1}{4} e_{pqk} e_{smn} n_p \left( \mathbb{B}_{rlts} n_r n_l \left( n_t D_k D_m + n_m D_k D_t \right) - \mathbb{B}_{rkts} \left( n_t D_r D_m + n_r D_m D_t + n_m D_t D_r \right) \right), \quad (76)$$

with $n_q H^{(1)}_{qj} = n_q \tilde{H}^{(1)}_{qj} = n_q H^{(2)}_{qj} = 0$. The differential system (73) is underdetermined since it comprises of 5 equations with 9 unknowns, namely, the components of the discontinuity vectors $g^{(p)}_j$ (with $p, j = 1, 2, 3$). It is worth noting that in the classical elasticity case, the differential system (73) degenerates into an *algebraic* system for the jump in the deformation gradient, providing the known condition of failure of ellipticity, $A^{(\mathbb{C})}_{qn} g^{(1)}_n = 0$ (Hill, 1961).

To further investigate the discontinuity relations, it is expedient to decompose the discontinuity vectors into their normal and tangential parts on the singular surface $\Sigma$ as

$$g^{(p)}_n = g^{\perp(p)}_n + g^{\|(p)}_n, \quad (77)$$

with

$$g^{\perp(p)}_n = \left( g^{(p)}_k n_k \right) n_n, \quad g^{\|(p)}_n = \left( \delta_{nq} - n_q n_n \right) g^{(p)}_q. \quad (78)$$

Accordingly, by taking into account Eqs. (37)$_2$, (74) and (75), it can be readily shown that the following relations hold true for the discontinuity vectors

$$A^{(\mathbb{B})}_{qn} g^{\perp(p)}_n = 0, \quad \tilde{H}^{(1)}_{qn} g^{\perp(p)}_n = -A^{(\mathbb{B})}_{qn} D_n \left( g^{(p)}_k n_k \right), \quad (79)$$

so that the differential system (73) becomes

$$\begin{cases} \left[ A_{qn}^{(\mathrm{C})} + H_{qn}^{(2)} \right] g_n^{(1)} - A_{qn}^{(\mathrm{B})} t_n^{\|} - \tilde{H}_{qn}^{(1)} g_n^{\|(2)} = 0 \\ A_{qn}^{(\mathrm{B})} g_n^{\|(2)} + H_{qn}^{(1)} g_n^{(1)} = 0 \end{cases}, \tag{80}$$

where $t_n^{\|} = g_n^{\|(3)} - D_n \left( g_k^{(2)} n_k \right)$ is a tangential vector on the surface $\Sigma$.

*The case of loss of ellipticity*

An interesting situation occurs at failure of (E) in the couple-stress medium. Consider for simplicity an intrinsic Cartesian coordinate system where the normal $\mathbf{n}$ at a given point on $\Sigma$ coincides with the local $x_1$-axis, so that $\mathbf{g}^{\|(p)} = \left( 0, g_2^{(p)}, g_3^{(p)} \right)$ and $\mathbf{t}^{\|} = \left( 0, t_2, t_3 \right)$. Assume further that $\mathbf{g}^{\|(2)}$ is an eigenvector of $\mathbf{A}^{(\mathrm{B})}(\mathbf{n})$ corresponding to a *null* eigenvalue (say $\lambda_2 = 0$ and $\lambda_3 \neq 0$), so that according to Eq. (63), (E) is lost and

$$\mathbf{A}^{(\mathrm{B})} \mathbf{g}^{\|(2)} = \mathbf{0}, \tag{81}$$

with $\mathbf{g}^{\|(2)} = \left( 0, g_2^{(2)}, 0 \right)$ being the pertinent eigenvector. Note that in the particular coordinate system chosen here, $A_{33}^{(\mathrm{B})} \equiv \lambda_3$ is the only non-vanishing component of $\mathbf{A}^{(\mathrm{B})}$, whereas, according to Eq. (81), $g_3^{(2)} = 0$.

In view of the above, the differential system (73) assumes the following explicit form

$$\begin{cases} A_{1j}^{(\mathrm{C})} g_j^{(1)} = 0 \\ \left[ A_{2j}^{(\mathrm{C})} + H_{2j}^{(2)} \right] g_j^{(1)} - \tilde{H}_{22}^{(1)} g_2^{(2)} = 0 \\ \left[ A_{3j}^{(\mathrm{C})} + H_{3j}^{(2)} \right] g_j^{(1)} - \lambda_3 t_3 - \tilde{H}_{32}^{(1)} g_2^{(2)} = 0, \quad j = 1, 2, 3, \\ H_{2j}^{(1)} g_j^{(1)} = 0 \\ H_{3j}^{(1)} g_j^{(1)} = 0 \end{cases} \tag{82}$$

with $t_3 = g_3^{(3)} - \partial_3 g_1^{(2)}$. The differential system comprises now of 5 unknowns, namely $g_1^{(1)}$, $g_2^{(1)}$, $g_3^{(1)}$, $g_2^{(2)}$, $t_3$, and 5 equations.

Therefore: *at loss of ellipticity the differential system becomes determinate*. The fact that the system becomes determinate proves that *discontinuous solutions can be possible at loss of ellipticity*. This statement will be validated with precise examples in Part II of this study, where it will be shown that at ellipticity loss folding of a continuum (in which the displacement gradient becomes discontinuous but remains finite) occurs under a concentrated force, and faulting (in which the displacement exhibits a finite discontinuity) occurs under a concentrated moment.

## 7. Antiplane strain

The general equations for an anisotropic couple-stress material and the relevant notions of ellipticity (E) and strong ellipticity (SE) are particularized in the present Section to the case of an orthotropic material under antiplane strain conditions. The case of antiplane strain is chosen for its comparative simplicity and since localization phenomena can be captured such as stress channelling, but also folding, and faulting.

### 7.1 Basic equations for an orthotropic material

For a body occupying a region in the $(x, y)$-plane, when antiplane strain conditions prevail, the displacement field reduces to

$$u_1 \equiv 0, \quad u_2 \equiv 0, \quad u_3 \equiv w(x, y) \neq 0. \tag{83}$$

Note that in Eq. (83) and henceforth, the coordinates $(x_1, x_2, x_3)$ are replaced by $(x, y, z)$, respectively, so that $z$ denotes the axis corresponding to the out-of-plane direction. By virtue of Eq. (10), the non-vanishing components of strain, rotation, and curvature are given in the form

$$\varepsilon_{xz} = \frac{1}{2}\frac{\partial w}{\partial x}, \quad \varepsilon_{yz} = \frac{1}{2}\frac{\partial w}{\partial y}, \quad \omega_x = \frac{1}{2}\frac{\partial w}{\partial y}, \quad \omega_y = -\frac{1}{2}\frac{\partial w}{\partial x},$$

$$\kappa_{xx} = -\kappa_{yy} = \frac{1}{2}\frac{\partial^2 w}{\partial x \partial y}, \quad \kappa_{xy} = -\frac{1}{2}\frac{\partial^2 w}{\partial x^2}, \quad \kappa_{yx} = \frac{1}{2}\frac{\partial^2 w}{\partial y^2}. \tag{84}$$

Further, considering an orthotropic centrosymmetric material, the constitutive equations (15) reduce to (for details refer to Appendix A)

$$\tau_{xz} = c_{55} \frac{\partial w}{\partial x}, \quad \tau_{yz} = c_{44} \frac{\partial w}{\partial y}, \tag{85}$$

$$m_{xx} = b_1 \kappa_{xx} = \frac{b_1}{2} \frac{\partial^2 w}{\partial x \partial y}, \quad m_{yy} = b_1 \kappa_{yy} = -\frac{b_1}{2} \frac{\partial^2 w}{\partial x \partial y},$$

$$m_{xy} = b_2 \kappa_{xy} + b_3 \kappa_{yx} = -\frac{b_2}{2} \frac{\partial^2 w}{\partial x^2} + \frac{b_3}{2} \frac{\partial^2 w}{\partial y^2},$$

$$m_{yx} = b_3 \kappa_{xy} + b_4 \kappa_{yx} = -\frac{b_3}{2} \frac{\partial^2 w}{\partial x^2} + \frac{b_4}{2} \frac{\partial^2 w}{\partial y^2}, \tag{86}$$

while, taking into account Eqs. (6) and (8), the (asymmetric) shear stresses become

$$\sigma_{xz} = c_{55} \frac{\partial w}{\partial x} - \frac{1}{4}\left( b_2 \frac{\partial^3 w}{\partial x^3} + (b_1 - b_3) \frac{\partial^3 w}{\partial x \partial y^2} \right), \tag{87}$$

$$\sigma_{yz} = c_{44} \frac{\partial w}{\partial y} - \frac{1}{4}\left( b_4 \frac{\partial^3 w}{\partial y^3} + (b_1 - b_3) \frac{\partial^3 w}{\partial x^2 \partial y} \right), \tag{88}$$

where $c_{44}$ and $c_{55}$ are the classical shear moduli characterizing an orthotropic Cauchy material subject to antiplane conditions, and $b_i$ ($i = 1,...,4$) are the couple-stress orthotropic moduli, with the dimension of a 'force'. In analogy with the theory of orthotropic Kirchhoff plates, the couple stress components $m_{xy}$ and $m_{yx}$ may be identified with the bending moments, while $m_{xx}$ and $m_{yy}$ with the twisting moments, applied on an element of a plate. In this context, the couple-stress parameters $b_2/2$ and $b_4/2$ represent the bending stiffnesses in the principal $y$- and $x$- directions, $b_1/2$ the principal twisting stiffness, and $b_3/2$ the stiffness associated with the effects of secondary bending (Lekhnitskii, 1963). Moreover, assuming the strain energy density $W$ to be positive definite (PD), the material moduli must satisfy the following inequalities

$$(\text{PD})^{\mathbb{C}} \Leftrightarrow c_{55} > 0, \quad c_{44} > 0, \tag{89}$$

$$(\text{PD})^{\mathbb{B}} \Leftrightarrow b_1 > 0, \quad b_2 > 0, \quad b_4 > 0, \quad b_2 b_4 - b_3^2 > 0. \tag{90}$$

Enforcing equilibrium yields a single PDE of the fourth-order for the out-of-plane displacement component

$$Lw + X_z + \frac{1}{2}\left(\frac{\partial Y_y}{\partial x} - \frac{\partial Y_x}{\partial y}\right) = 0, \qquad (91)$$

where the differential operator $L$ is defined as

$$L \equiv \underbrace{c_{55}\partial_x^2 + c_{44}\partial_y^2}_{L^{cl}\,\equiv\,\text{lower (classical) part}} - \frac{1}{4}\underbrace{\left(b_2\partial_x^4 + 2b_0\partial_x^2\partial_y^2 + b_4\partial_y^4\right)}_{L^{cs}\,\equiv\,\text{principal part}}, \qquad (92)$$

and $b_0 = b_1 - b_3$ is a material parameter that accounts for both torsion and secondary bending effects. Note that neglecting the body forces and body moments, the equilibrium equation (91) is of the same form as the equation of bending of thin orthotropic plates with *prestress* (Lekhnitskii, 1963). In particular, the principal part $L^{cs} w$ is associated with the deflection of the plate, whereas the classical (lower) part $L^{cl} w$ plays the role of the prestress.

Finally, it is remarked that when $c_{44} = c_{55} = \mu$, $b_1 = 4\eta + 4\eta'$, $b_2 = b_4 = 4\eta$, and $b_3 = 4\eta'$, the above equations transform to those governing *isotropic* couple-stress elasticity for antiplane strain deformations (Lubarda, 2003). In that case, the scalar equation of equilibrium (91) reduces to

$$\mu\nabla^2 w - \eta\nabla^4 w + X_z + \frac{1}{2}\left(\frac{\partial Y_y}{\partial x} - \frac{\partial Y_x}{\partial y}\right) = 0. \qquad (93)$$

*7.2 Ellipticity*

It is well-known for classical Cauchy materials that loss of ellipticity may result in the emergence of various kinds of discontinuities. Therefore, since conditions of stress channelling and associated localized solutions are addressed in the present study, it is expedient to classify the governing PDE in Eq. (91) with respect to its elliptic regime. For the antiplane strain case considered here, and in the light of the results obtained in Section 5, the (E) condition becomes

$$\lambda_3(\mathbf{n}) = b_2 n_x^4 + 2b_0 n_x^2 n_y^2 + b_4 n_y^4 \neq 0, \quad \forall \mathbf{n}: |\mathbf{n}| = 1, \tag{94}$$

where $\lambda_3(\mathbf{n})$ is the relevant eigenvalue in antiplane strain of the couple-stress part $\mathbf{A}^{(\mathbb{B})}$ of the acoustic tensor, Eq. (63)$_3$.

To investigate the restrictions that condition (94) imposes on the couple-stress moduli, the *principal* operator $L^{cs}$ of the governing PDE (91) is examined (Renardy and Rogers, 2004). In particular, a solution to the equation $L^{cs} w = 0$ is assumed in the form (Bigoni, 2012)

$$w = F(x + \Psi y), \tag{95}$$

where $F$ is an analytical function and $\Psi$ is a complex constant satisfying the following bi-quadratic equation, obtained by inserting representation (95) in the principal part of Eq. (91)

$$\Psi^4 + 2\gamma \Psi^2 + \beta = 0, \tag{96}$$

with $\beta = b_2/b_4$ being the ratio between the principal bending moduli, and $\gamma = b_0/b_4$. The four-roots $\Psi_q$ ($q = 1,...,4$) of Eq. (96) satisfy

$$\Psi_q^2 = -\gamma + (-1)^q \sqrt{\gamma^2 - \beta}, \tag{97}$$

which are real or complex depending on the values of $b_2$, $b_4$ and $b_0$, or equivalently the dimensionless parameters $\beta$ and $\gamma$. In what follows, unless otherwise stated, it is assumed that $b_4 > 0$.

The roots $\Psi_q$ defined by Eq. (97) change their nature according to the values taken by parameters $\beta$ and $\gamma$, so that they can be classified as follows.

*Elliptic imaginary regime (EI)*
If $\beta > 0$ and $\gamma \geq \sqrt{\beta}$ (or equivalently $b_2 > 0$ and $b_0 \geq \sqrt{b_2 b_4}$), Eq. (96) admits four purely imaginary roots

$$\Psi_1 = ic_1, \quad \Psi_2 = ic_2, \quad \Psi_3 = \bar{\Psi}_1, \quad \Psi_4 = \bar{\Psi}_2, \tag{98}$$

with

$$\left.\begin{array}{l} c_1 \\ c_2 \end{array}\right\} = \sqrt{\gamma \pm \sqrt{\gamma^2 - \beta}} > 0. \tag{99}$$

***Elliptic complex regime (EC)***

If $\beta > 0$ and $-\sqrt{\beta} < \gamma < \sqrt{\beta}$ (or equivalently $b_2 > 0$ and $-\sqrt{b_2 b_4} < b_0 < \sqrt{b_2 b_4}$), Eq. (96) admits four complex conjugate roots

$$\Psi_1 = -f + ic, \quad \Psi_2 = f + ic, \quad \Psi_3 = \bar{\Psi}_1, \quad \Psi_4 = \bar{\Psi}_2, \tag{100}$$

with

$$\left.\begin{array}{l} f \\ c \end{array}\right\} = \sqrt{\frac{\sqrt{\beta} \mp \gamma}{2}} > 0. \tag{101}$$

***Hyperbolic regime (H)***

If $\beta > 0$ and $\gamma \leq -\sqrt{\beta}$ (or equivalently $b_2 > 0$ and $b_0 \leq -\sqrt{b_2 b_4}$), Eq. (96) admits four real roots

$$\Psi_{1,3} = \pm e_1, \quad \Psi_{2,4} = \pm e_2, \tag{102}$$

with

$$\left.\begin{array}{l} e_1 \\ e_2 \end{array}\right\} = \sqrt{-\gamma \pm \sqrt{\gamma^2 - \beta}} > 0. \tag{103}$$

***Parabolic regime (P)***

If $\beta \leq 0$, Eq. (96), admits two real and two imaginary roots, namely

$$\Psi_1 = f, \quad \Psi_2 = ic, \quad \Psi_3 = -\Psi_1, \quad \Psi_4 = -\Psi_2, \tag{104}$$

with

$$\left.\begin{array}{c} f \\ c \end{array}\right\} = \sqrt{\mp\gamma + \sqrt{\gamma^2 - \beta}} > 0. \tag{105}$$

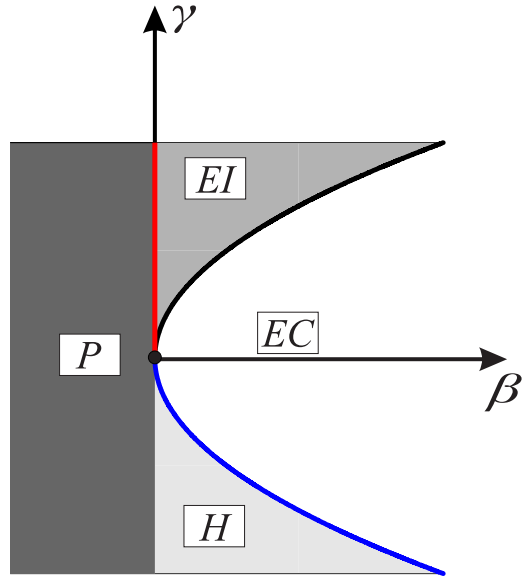

**Fig. 3**: Regime classification in the $\beta = b_2/b_4$ versus $\gamma = b_0/b_4$ parameter space for an orthotropic couple-stress material under antiplane strain conditions. The (EI/P) and (EC/H) ellipticity boundaries are depicted with red and blue lines, respectively.

As a conclusion, the (E) condition in a couple-stress material under antiplane strain deformation assumes the following form

$$(\text{E}) \Leftrightarrow b_2 > 0 \text{ and } b_0 > -\sqrt{b_2 b_4}, \tag{106}$$

holding for $b_4 > 0$. Equivalently, the ellipticity conditions can be written as $\beta > 0$ and $\gamma > -\sqrt{\beta}$ (Fig. 3). Therefore, ellipticity (E) can be lost in two ways: either (i.) at the (EI/P) boundary (red line in Fig. 3), where $\beta = 0$ and $\gamma > 0$ (i.e. $b_2 = 0$ and $b_0 > 0$), or (ii.) at the (EC/H) boundary (blue line in Fig. 3), where $\beta > 0$ and $\gamma = -\sqrt{\beta}$ (i.e. $b_2 b_4 > 0$ and $b_0 = -\sqrt{b_2 b_4}$). In particular, at the (EI/P) boundary only one possible discontinuity surface emerges which is aligned parallel to the $y$-axis. On the other hand, at the (EC/H) boundary two discontinuity surfaces are possible. The inclination angle $\varphi$ (with the $x$-axis) of the normal to the discontinuity surfaces depends solely upon the ratio $\beta$, and can be calculated at the (EC/H) boundary using Eq. (94), as

$$\tan^2 \varphi = \sqrt{\frac{b_2}{b_4}} = \sqrt{\beta}, \tag{107}$$

corresponding to the following components of the unit vector $\mathbf{n}$ normal to the discontinuity surfaces

$$n_x = \pm \sqrt{\frac{\sqrt{b_4}}{\sqrt{b_2} + \sqrt{b_4}}}, \quad n_y = \pm \sqrt{\frac{\sqrt{b_2}}{\sqrt{b_2} + \sqrt{b_4}}}. \tag{108}$$

The implications of both cases of ellipticity loss on the behavior of the solution and the emergence of the associated discontinuities in the components of the displacement and its gradient will be examined in Part II. As a conclusion from the above-discussion, note that loss of uniqueness, as related to failure of (PD), for a problem of antiplane deformation can be lost simultaneously with (E). Therefore, a material can be designed to work in antiplane conditions and display extreme behaviors (such as stress channelling and emergence of localized folding and faulting), but still preserving uniqueness of the solution and wave propagation.

*7.3 SH waves and the (WP) condition*

Antiplane shear (i.e. horizontally polarized or SH) motions are now examined in a homogeneous orthotropic constrained Cosserat medium. To this purpose, the governing equation (91) is augmented with the classical inertia term $\rho \ddot{w}$. Assuming zero body forces and moments, the equation of motion for the out-of-plane displacement becomes

$$Lw = \rho \ddot{w}. \tag{109}$$

Substituting into the equation of motion a plane wave harmonic solution of the form

$$w(x, y) = d_3 e^{-ik(\mathbf{x} \cdot \mathbf{n} - V_S t)}, \tag{110}$$

the dispersion equation is obtained relating the phase velocity $V_S$ of SH waves to the wavenumber $k$

$$V_S^2 = \rho^{-1}\left[c_{55}n_x^2 + c_{44}n_y^2 + \frac{k^2}{4}\left(b_2 n_x^4 + 2b_0 n_x^2 n_y^2 + b_4 n_y^4\right)\right]. \tag{111}$$

For an orthotropic material under antiplane motions, the only non-vanishing out-of-plane component of the acoustic tensor is

$$A_{33}(k, \mathbf{n}) = \rho k^2 V_S^2. \tag{112}$$

The (WP) condition requires that SH waves propagate with real non-zero velocities for all real wavenumbers $k$ in every direction $\mathbf{n}$, which, in turn, implies that $A_{33}(k, \mathbf{n}) > 0$. In view of the inequalities stated in Eq. (40), the (WP) condition can be defined for antiplane deformations as

$$(\text{WP}) \Leftrightarrow c_{55}n_x^2 + c_{44}n_y^2 \geq 0, \quad b_2 n_x^4 + 2b_0 n_x^2 n_y^2 + b_4 n_y^4 \geq 0, \quad \text{and} \quad A_{33} \neq 0. \tag{113}$$

The conditions for the first inequality in Eq. (113) to hold require that $c_{55} \geq 0$ and $c_{44} \geq 0$, whereas from the second inequality it can be deduced (c.f. Eqs. (106), Section 7.2) that $b_2 \geq 0$, and $b_0 \geq -\sqrt{b_2 b_4}$. It is worth noting that when $c_{55} = c_{44} = 0$, the propagation velocity of SH waves depends only upon the Cosserat moduli. Consequently, the propagation velocity becomes a *linear* function of the wave number $k$, a situation analogous to the classical plate theory (Graff, 1975). Moreover, it is remarked that in the special case in which

$$\lambda_3(\mathbf{n}) = b_2 n_x^4 + 2b_0 n_x^2 n_y^2 + b_4 n_y^4 = 0, \tag{114}$$

ellipticity is lost, but waves may still propagate. For example, assuming that $b_2 = 0$, $b_0 > 0$, and $b_4 > 0$, the condition of (E) fails since in the direction $\mathbf{n} = (\pm 1, 0)$, $\lambda_3(\mathbf{n})$ becomes zero. However, for $c_{55} > 0$ and $c_{44} \geq 0$, SH waves can still propagate for *all* wavenumbers and directions of propagation $\mathbf{n}$. *This shows that (WP) does not imply (E).*

Finally, the (SE) conditions for the elasticity tensors in the antiplane strain case are obtained from the general inequalities in Eqs. (18), with $\mathbf{q}$ parallel to the $z$-axis and $\mathbf{n}$ orthogonal. In particular, it can be shown that (SE) requires that

$$(\text{SE})^\mathbb{C} \Leftrightarrow c_{55} > 0, \ c_{44} > 0, \tag{115}$$

$$(\text{SE})^\mathbb{B} \Leftrightarrow b_1 > 0, \ b_2 > 0, \ b_4 > 0, \ |b_3| < b_1 + \sqrt{b_2 b_4}. \tag{116}$$

It follows, as expected, that (SE) implies both the (WP) and (E) conditions. In the classical elasticity case, the notions of (SE) and (WP) coalesce and the pertinent conditions become $c_{55} > 0$ and $c_{44} > 0$.

## 8. Extreme Cosserat materials

An extreme Cosserat material is defined in a way that its mechanical properties are close to ellipticity loss. An interesting case occurs when both the response of the Cosserat material and of the underlying Cauchy solid (obtained setting the Cosserat stiffness to zero) are simultaneously close to a failure of ellipticity. In all these cases, the extreme material is on the verge of a material instability, but still within the elliptic range. Although these extreme materials will be chosen so that waves can still propagate –the (WP) condition–, it will be shown through specific antiplane strain examples (deferred to Part II of this study) that stress channelling is related to the 'distance' to failure of ellipticity. In this context, localized folding and faulting will also emerge.

In addition to the above-mentioned cases, it is possible to figure out the existence of extreme Cosserat materials, in which some Cauchy elasticity stiffness vanishes. Examples of these materials are sketched in Fig. 4 where: (i.) a 'frictionless telescopic tube' is a structure having null longitudinal elastic modulus, but still able to carry a variable bending moment and a shear force and (ii.) a jointed beam (in which all segments are connected with frictionless joints) has null transversal shear modulus, but is still able to carry a longitudinal force and a constant bending moment. Examples of these kinds of materials are specified below.

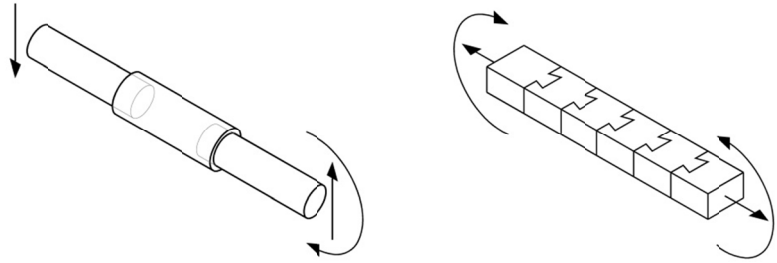

**Fig. 4** Examples of 'extreme materials' in which a Cauchy elastic stiffness is null, but Cosserat effects exist. Left: a frictionless telescopic tube is an element with null longitudinal elasticity, but able to carry a variable moment and related shear force. Right: a jointed beam (with perfectly smooth joints) has null transversal shear modulus, but can carry a longitudinal force and a constant moment.

*8.1 A purely Cosserat solid*

A purely Cosserat solid is defined as a material with null Cauchy stiffness, $\mathbb{C} = \mathbf{0}$. For this material the stress is only antisymmetric, so that couples and shear stresses are generated (even if the shear stiffness is null)

$$\tau_{pq} = 0, \qquad m_{pq} = \mathbb{B}_{pqmn} \kappa_{mn}, \qquad \alpha_{pq} = -\frac{1}{2} e_{pqk} m_{rk,r}. \qquad (117)$$

This material is not elliptic in a 3D framework, but ellipticity can be restored in a framework allowing only for antiplane deformation. In this framework, (PD) is lost (as related to the Cauchy part of the constitutive equation), but SH-waves still propagate with finite speed.

*8.2 A shear-defective Cauchy material cured with a strongly elliptic Cosserat part*

A shear-defective Cauchy material 'cured' with a strongly elliptic Cosserat part is defined as a material with isotropic Cauchy elasticity with null shear stiffness, augmented with an elliptic Cosserat part.

$$\tau_{pq} = \lambda \varepsilon_{kk} \delta_{pq}, \quad m_{pq} = \mathbb{B}_{pqmn} \kappa_{mn}, \quad \alpha_{pq} = -\frac{1}{2} e_{pqk} m_{rk,r}. \tag{118}$$

For this material (PD) is lost (as related to the loss of shear stiffness), but ellipticity is preserved and waves can propagate with positive speed. For instance, assuming that $\mathbb{B}$ is isotropic, the phase velocities of a longitudinal and shear waves become, respectively

$$V_P = \sqrt{\lambda/\rho}, \quad V_S = k\sqrt{\eta/\rho}, \tag{119}$$

where $V_S$ coincides with the speed of flexural waves in a beam (Graff, 1975). Note, however, that if the shear modulus is negative ($\mu < 0$), waves cannot propagate, but for $\eta > 0$ the material is still in the elliptic range, Eq. (64)$_2$. This proves that *(E) does not imply (WP)*, so that recalling an observation made in the previous Section, we conclude that: *ellipticity (E) and wave propagation (WP) are not interdependent conditions*.

## 9. Conclusions

The definition of anisotropic extreme Cosserat materials and their material stability analysis has led to several results defining the behavior of these materials, which can be listed as follows.

i.) Positive definiteness of the strain energy (PD) is related to both (PD)$^\mathbb{C}$ of the Cauchy and (PD)$^\mathbb{B}$ of the Cosserat parts of the elasticity, so that loss of (PD) for a Cauchy material cannot be restored by augmenting the constitutive equation with a Cosserat part.

ii.) A notion of strong ellipticity (SE) has been introduced (reducing to the 'usual' notion for classical Cauchy elasticity), which implies uniqueness for a problem with kinematics prescribed on the whole boundary of a homogeneous solid, that has been proven through an extension of the van Hove theorem.

iii.) Strong ellipticity (SE) implies that planar waves can propagate with real speed (WP) and that the partial differential operator governing equilibrium is elliptic (E). A necessary condition for (WP) is that the Cauchy part of elasticity allows propagation of pressure waves.

iv.) Ellipticity (E) and wave propagation (WP) condition are not interdependent conditions, so that it can happen that a wave cannot propagate when the material is still in the elliptic range, or vice versa that waves propagate when the material is at the boundary of ellipticity loss.

v.) Investigation of emergence of discontinuity surfaces in Cosserat materials showed that failure of (E) is expected to give rise to the emergency of localized deformations (as will be demonstrated with specific examples of localized folding and faulting of a continuum in Part II of this study).

*Acknowledgements* - Financial support from the ERC advanced grant 'Instabilities and nonlocal multiscale modelling of materials' FP7-PEOPLE-IDEAS-ERC-2013-AdG (2014-2019) is gratefully acknowledged.

## Appendix A. Elastic moduli for orthotropic couple-stress materials

For an orthotropic material the couple-stress elasticity tensor $\mathbb{B}_{ijkl}$, possessing the major symmetry and subject to the restriction $\mathbb{B}_{ijkk} = \mathbb{B}_{kkij} = \mathbf{0}$, transforms according to the relation

$$\mathbb{B}_{ijkl} = R_{ip} R_{jq} R_{kr} R_{ls} \mathbb{B}_{pqrs} \tag{A1}$$

where $R_{ip}$ are the orthogonal tensors that characterize the three mutually perpendicular planes of reflective symmetry (Cowin, 2013). In particular, it can be shown that the couple-stress elasticity tensor $\mathbb{B}$ has 12 independent components in the orthotropic case and can be represented in a matrix form in the following way

$$\mathbb{B} = \begin{bmatrix} \mathbb{B}_{1111} & 0 & 0 & 0 & \mathbb{B}_{1122} & 0 & 0 & 0 & \mathbb{B}_{1133}^* \\ & \mathbb{B}_{1212} & 0 & \mathbb{B}_{1221} & 0 & 0 & 0 & 0 & 0 \\ & & \mathbb{B}_{1313} & 0 & 0 & 0 & \mathbb{B}_{1331} & 0 & 0 \\ & & & \mathbb{B}_{2121} & 0 & 0 & 0 & 0 & 0 \\ & & & & \mathbb{B}_{2222} & 0 & 0 & 0 & \mathbb{B}_{2233}^* \\ & \text{SYM} & & & & \mathbb{B}_{2323} & 0 & \mathbb{B}_{2332} & 0 \\ & & & & & & \mathbb{B}_{3131} & 0 & 0 \\ & & & & & & & \mathbb{B}_{3232} & 0 \\ & & & & & & & & \mathbb{B}_{3333}^* \end{bmatrix}, \quad (A2)$$

when curvatures and couple stresses are represented as the following vectors

$$\boldsymbol{\kappa} = \{\kappa_{11}, \kappa_{12}, \kappa_{13}, \kappa_{21}, \kappa_{22}, \kappa_{23}, \kappa_{31}, \kappa_{32}, \kappa_{33}\}, \quad (A3)$$

$$\mathbf{m} = \{m_{11}, m_{12}, m_{13}, m_{21}, m_{22}, m_{23}, m_{31}, m_{32}, m_{33}\}. \quad (A4)$$

Note that some components in Eq. (A2) are not independent, so that

$$\mathbb{B}_{1133}^* = -(\mathbb{B}_{1111} + \mathbb{B}_{1122}), \quad \mathbb{B}_{2233}^* = -(\mathbb{B}_{1122} + \mathbb{B}_{2222}), \quad \mathbb{B}_{3333}^* = -(\mathbb{B}_{1133}^* + \mathbb{B}_{2233}^*). \quad (A5)$$

The constitutive equations follow from Eq. (15)$_2$, where it can be easily verified that $m_{pp} = 0$. Note that for micropolar elasticity (the so-called 'unconstrained Cosserat theory'), not considered here, an orthotropic material is characterized by 15 independent components since $\mathbb{B}_{ijkk} \neq 0$ (Ilcewicz et al., 1986).

In the special case of antiplane strain, all the in-plane degrees of freedom are zero, and thus: $u_x = u_y = \omega_z = 0$. According to Eq. (84), the non-vanishing components of the curvature tensor are $(\kappa_{xx}, \kappa_{xy}, \kappa_{yx}, \kappa_{yy})$. Note, that $\kappa_{pp} = 0$ and $\kappa_{zz} = 0$, so that $\kappa_{xx} = -\kappa_{yy}$, in this particular case. Accordingly, the constitutive equations for an orthotropic couple-stress material under antiplane conditions become

$$m_{xx} = \mathbb{B}_{1111}\kappa_{xx} + \mathbb{B}_{1122}\kappa_{yy}, \quad m_{xy} = \mathbb{B}_{1212}\kappa_{xy} + \mathbb{B}_{1221}\kappa_{yx},$$
$$m_{yx} = \mathbb{B}_{1221}\kappa_{xy} + \mathbb{B}_{2121}\kappa_{yx}, \quad m_{yy} = \mathbb{B}_{1122}\kappa_{xx} + \mathbb{B}_{2222}\kappa_{yy}. \quad (A6)$$

Therefore, a couple-stress material in antiplane strain is governed by 6 microstructural constants, additional to the 2 classical shear moduli $c_{44}$ and $c_{55}$. By noting that $\kappa_{xx} = -\kappa_{yy}$, and assuming, for simplicity that the couple-stress material possess the same principal torsional stiffness in the $x$- and $y$- directions and *null* secondary torsional stiffness $\mathbb{B}_{1122} = 0$, the constitutive equations reduce to (Section 7.1)

$$m_{xx} = b_1 \kappa_{xx}, \quad m_{yy} = b_1 \kappa_{yy}, \quad m_{xy} = b_2 \kappa_{xy} + b_3 \kappa_{yx}, \quad m_{yx} = b_3 \kappa_{xy} + b_4 \kappa_{yx}, \tag{A7}$$

with $b_1 \equiv \mathbb{B}_{1111} = \mathbb{B}_{2222}$, $b_2 \equiv \mathbb{B}_{1212}$, $b_3 \equiv \mathbb{B}_{1221}$, and $b_4 \equiv \mathbb{B}_{2121}$.

## Appendix B. Two alternative derivations of the ellipticity condition

An alternative way to obtain the ellipticity condition for the degenerate couple-stress operator in Eq. (53) is to examine the behavior of the determinant of the total symbol (54) at large $k$. In general, if the principal operator is not degenerate, the behavior of the total symbol matrix at large $k$ provides the same condition obtained by examining the principal symbol. However, when the principal operator is degenerate, the determinant of the total symbol at large $k$ involves also terms associated with the lower-order operator. In the present case, the determinant of the total symbol $l$ degenerates to a polynomial of the tenth-degree with respect to the modulus $k$, exhibiting the following asymptotic behavior as $k \to \infty$

$$\det l = \det \mathbf{A} = (\tau_\nu \lambda_2 \lambda_3) k^{10} \quad \text{as} \quad k \to \infty, \tag{B1}$$

where $\tau_\nu(\mathbf{n}) = \mathbf{n} \cdot \mathbf{A}^{(C)} \mathbf{n}$, and $\lambda_2 \equiv \lambda_2(\mathbf{n})$, $\lambda_3 \equiv \lambda_3(\mathbf{n})$ are the non-vanishing eigenvalues of $\mathbf{A}^{(\mathbb{B})}$. Requiring that $\det l \neq 0$, the conditions of ellipticity stated in Eqs. (63) are recovered.

A second, more rigorous, approach to derive the condition of ellipticity is now developed. The approach is based on the Douglis-Nirenberg definition of ellipticity (Douglis and Nirenberg, 1955; Agmon et al., 1959), which is more general than the standard definition.

First, the governing system is transformed into an equivalent higher-order system by taking the divergence and the curl of Eqs. (52), respectively

$$\text{div}\left(\mathbf{L}^{(C)}\mathbf{u}\right)=0, \quad \nabla_n\times\left(\mathbf{L}\mathbf{u}\right)=\mathbf{0}, \tag{B2}$$

where $\nabla_n=\mathbf{n}D$, $D=n_q\partial_q$ is the directional derivative taken in the direction of the unit vector $\mathbf{n}$, and Eq. (56) has been used.

The scalar Eq. (B2)$_1$ is a third-order PDE and the vectorial equation (B1)$_2$ is a system of PDEs of the fifth-order. Employing an orthonormal basis $(\mathbf{n},\mathbf{t},\mathbf{s})$, and noting that $\mathbf{n}\cdot\nabla_n\times(\mathbf{L}\mathbf{u})\equiv 0$, the projections of $\nabla_n\times(\mathbf{L}\mathbf{u})$ in the plane spanned by the vectors $\mathbf{t}$ and $\mathbf{s}$ assume the form

$$\mathbf{t}\cdot\left[\nabla_n\times(\mathbf{L}\mathbf{u})\right]=0, \quad \mathbf{s}\cdot\left[\nabla_n\times(\mathbf{L}\mathbf{u})\right]=0. \tag{B3}$$

Further, by noting that

$$e_{ijq}t_i n_j L_{qn}Du_n=-s_q L_{qn}Du_n, \quad e_{ijq}s_i n_j L_{qn}Du_n=t_q L_{qn}Du_n, \tag{B4}$$

equations (B3) become

$$\mathbf{t}\cdot\mathbf{L}(D\mathbf{u})=0, \quad \mathbf{s}\cdot\mathbf{L}(D\mathbf{u})=0, \tag{B5}$$

so that the original governing system of equations (52) assumes the equivalent form

$$\begin{cases} \text{div}\left(\mathbf{L}^{(C)}\mathbf{u}\right)=0 \\ \mathbf{t}\cdot\mathbf{L}^{(C)}(D\mathbf{u})+\mathbf{t}\cdot\mathbf{L}^{(B)}(D\mathbf{u})=0, \\ \mathbf{s}\cdot\mathbf{L}^{(C)}(D\mathbf{u})+\mathbf{s}\cdot\mathbf{L}^{(B)}(D\mathbf{u})=0 \end{cases} \tag{B6}$$

The above system comprises of elements which have different orders, in such cases, ellipticity can be defined in the general Douglis-Nirenberg sense, where terms of different orders may be included in the principal part. Note that such an approach could not be employed in the original system (52), since *all* elements of the original operator $L_{qn}$ were of the fourth-order.

According to the Douglis-Nirenberg definition of ellipticity, a system is elliptic if there exist integer weights $\xi_\mu$ and $\zeta_\nu$ for which $\deg(L_{qn}) \leq \xi_\mu + \zeta_\nu$ and $\det(l^P_{qn}(i\mathbf{k})) \neq 0$ for all $\mathbf{k} \in \mathbb{R}^3$ with $|\mathbf{k}| \neq 0$. The principal part $l^P_{qn}$ consists of all those terms which have order equal to $\xi_\mu + \zeta_\nu$. Note that this definition is equivalent to the standard one only when all $\xi_\mu$ are equal.

For the system at hand in Eq. (B6), the weights $\xi_1 = 3$, $\xi_2 = \xi_3 = 5$, $\zeta_1 = \zeta_2 = \zeta_3 = 0$ are assigned. The principal operator becomes

$$\begin{cases} \operatorname{div}(\mathbf{L}^{(C)}\mathbf{u}) = 0 \\ \mathbf{t} \cdot \mathbf{L}^{(B)}(D\mathbf{u}) = 0 \\ \mathbf{s} \cdot \mathbf{L}^{(B)}(D\mathbf{u}) = 0 \end{cases}. \tag{B7}$$

Further, the displacement vector is represented as $\mathbf{u} = u_\nu \mathbf{n} + u_\tau \mathbf{t} + u_\sigma \mathbf{s}$ in the orthonormal basis $(\mathbf{n}, \mathbf{t}, \mathbf{s})$. The principal symbol matrix assumes now the following explicit form

$$l^P(i\mathbf{k}) = ik^3 \begin{bmatrix} \mathbf{n} \cdot \mathbf{A}^{(C)}\mathbf{n} & \mathbf{n} \cdot \mathbf{A}^{(C)}\mathbf{t} & \mathbf{n} \cdot \mathbf{A}^{(C)}\mathbf{s} \\ 0 & -k^2 \mathbf{t} \cdot \mathbf{A}^{(B)}\mathbf{t} & -k^2 \mathbf{t} \cdot \mathbf{A}^{(B)}\mathbf{s} \\ 0 & -k^2 \mathbf{s} \cdot \mathbf{A}^{(B)}\mathbf{t} & -k^2 \mathbf{s} \cdot \mathbf{A}^{(B)}\mathbf{s} \end{bmatrix}, \tag{B8}$$

where $\mathbf{k} = k\mathbf{n}$. Moreover, taking into account that $\mathbf{A}^{(B)}\mathbf{n} = \mathbf{0}$, the minor determinant $M_{11}$ of the above matrix can be equivalently written as $M_{11} = k^4 \lambda_2 \lambda_3$. Therefore, the determinant of the principal symbol finally becomes

$$\det l^P = -ik^{13} \tau_\nu \lambda_2 \lambda_3. \tag{B9}$$

The condition of ellipticity (E) for a constrained Cosserat material is that the determinant of the principal symbol is different from zero, which reduces to the same conditions provided by Eqs. (63) and Eqs. (B1). It can be therefore stated that the couple-stress operator is elliptic in the more general Douglis-Nirenberg sense.